\documentclass[preprint,3p,times]{elsarticle}

\usepackage{amssymb}
\usepackage{amsmath}

\usepackage{color}
\usepackage{hyperref}
\hypersetup{colorlinks, linkcolor=blue}

\journal{Optik}

\makeatletter
\def\ps@pprintTitle{%
  \let\@oddhead\@empty
  \let\@evenhead\@empty
  \let\@oddfoot\@empty
  \let\@evenfoot\@oddfoot
}
\makeatother

\begin{document}

\begin{frontmatter}



\title{(Invited) Two-color soliton meta-atoms and molecules}

\author[inst1,inst2]{O. Melchert}
\author[inst1,inst2]{S. Willms}
\author[inst1,inst2]{I. Babushkin}
\author[inst1,inst2]{U. Morgner}
\author[inst1,inst2]{A. Demircan}

\affiliation[inst1]{organization={Leibniz Universit\"at Hannover, Institute of Quantum Optics}, 
            addressline={Welfengarten 1}, 
            city={Hannover},
            postcode={30167}, 
            country={Germany}}

\affiliation[inst2]{organization={Leibniz Universit\"at Hannover, Cluster of Excellence PhoenixD}, 
            addressline={Welfengarten 1A}, 
            city={Hannover},
            postcode={30167}, 
            country={Germany}}

\begin{abstract}
We present a detailed overview of the physics of two-color soliton molecules in
nonlinear waveguides, i.e.\ bound states of localized optical pulses which are
held together due to an incoherent interaction mechanism. 
The mutual confinement, or trapping, of the subpulses, which leads to a stable
propagation of the pulse compound, is enabled by the nonlinear Kerr effect.
Special attention is paid to the description of the binding mechanism in terms
of attractive potential wells, induced by the refractive index changes of the
subpulses, exerted on one another through cross-phase modulation. 
Specifically, we discuss nonlinear-photonics meta atoms, given by pulse
compounds consisting of a strong trapping pulse and a weak trapped pulse, for
which trapped states of low intensity are determined by a Schrödinger-type
eigenproblem.
We discuss the rich dynamical behavior of such meta-atoms, demonstrating that
an increase of the group-velocity mismatch of both subpulses leads to an
ionization-like trapping-to-escape transition.
We further demonstrate that if both constituent pulses are of similar
amplitude, molecule-like bound-states are formed. 
We show that $z$-periodic amplitude variations permit a coupling of these pulse
compound to dispersive waves, resulting in the resonant emission of
Kushi-comb-like multi-frequency radiation.
\end{abstract}



\begin{keyword}
Nonlinear optics \sep Optical solitons \sep Two-color soliton molecules \sep Resonant radiation


\end{keyword}

\end{frontmatter}


\tableofcontents

\section{Introduction}
\label{sec:intro}

The confinement of two -- and possibly more -- quasi co-propagating optical
pulses has been discussed in terms of various propagation settings since the
80's of the last century, 
with early accounts discussing the self-confinement multimode optical pulses in
glass fibers \cite{Hasegawa:OL:1980}, nonlinear pairing of light and dark
optical solitons \cite{Afanasjev:JETP:1988,Afanasjev:JQE:1989}, and stability
of solitons with different polarization components in birefringent fibers
\cite{Menyuk:OL:1987}.
A very paradigmatic instance of self-confinement is supported by the standard
nonlinear Schrödinger equation (NSE)
\cite{Mitschke:BOOK:2016,Agrawal:BOOK:2019}. 
In the integrable case, it features localized field pulses given by solitary
waves \cite{Drazin:BOOK:1989}. 
When considering two or more quasi group-velocity matched pulses, their
incoherent, cross-phase modulation (XPM) induced mutual interaction
co-determines their dynamics
\cite{Hasegawa:OL:1980,Afanasjev:JETP:1988,Afanasjev:JQE:1989,Menyuk:OL:1987,Haelterman:PRE:1994,Mitchell:PRL:1997,Tan:CSF:2000,Tan:JAS:1995}. 
%
For instance, in nonlinear waveguides with a single zero-dispersion point, a
soliton induces a strong refractive index barrier that cannot be surpassed by
quasi group-velocity matched waves located in a domain of normal dispersion
\cite{Demircan:PRL:2011}, resulting in their mutual repulsion. 
The underlying interaction process is enabled by a general wave reflection
mechanism originally reported in fluid dynamics \cite{Smith:PCPS:1975}.  In
optics this process is referred to as push-broom effect
\cite{deSterke:OL:1992}, optical event horizon
\cite{Philbin:S:2008,Faccio:CP:2012}, or temporal reflection
\cite{Plansinis:PRL:2015}. 
This interaction mechanism allows for a strong and efficient control of light
pulses \cite{Demircan:PRL:2013,Demircan:OL:2014,Babushkin:IEEE:2016}, and has
been shown to appear naturally during the supercontinuum generation process
\cite{Driben:OE:2010,Demircan:APB:2014,Skryabin:RMP:2010}.
When considering waveguides that support group-velocity matched propagation of
pulses in separate domains of anomalous dispersion, their mutual interaction is
expressed in a different way:
the aforementioned XPM induces attractive potentials that hold the pulses
together, enabling two-color soliton molecules through an incoherent binding
mechanism \cite{Melchert:PRL:2019}; the resulting pulse compound consists of
two subpulses at vastly different center frequencies.
Putting emphasis on the frequency-domain representation of these pulse
compounds lead to observe that a soliton can in fact act as a localized
trapping potential with a discrete level spectrum \cite{Melchert:PRL:2019}.
Let us emphasize that in order to achieve a strong attractive interaction
between the subpulses of such pulse compounds, group-velocity matching is
crucial \cite{Melchert:SR:2021}.
In terms of a modified NSE with added fourth-order dispersion, these objects
where identified as parts of a large family of generalized dispersion Kerr
solitons that can be characterized using the concept of a meta-envelope
\cite{Tam:PRA:2020}.
Such pulses were recently verified experimentally in mode-locked
laser cavities \cite{Lourdesamy:NP:2021,Mao:NC:2021,Cui:PP:2022}. 
In a complementary approach to the multi-scales analysis presented in
Ref.~\cite{Tam:PRA:2020}, modeling both subpulses in terms of coupled NSEs
allowed to derive a special class of two-color soliton pairs and their
meta-envelopes in closed form \cite{Melchert:OL:2021}.
Let us note that the concept of soliton molecules has meanwhile been extended
to pulse compounds with three frequency centers \cite{Willms:PRA:2022}, and
recently also to a number of $J$ equally spaced frequency components
\cite{Lourdesamy:NP:2021,Lourdesamy:JOSAB:2023}.
Further, two-color soliton microcomb states with similar structure where also
observed in the framework of the Lugiato-Lefever equation
\cite{Melchert:OL:2020,Moille:OL:2018}.
The underlying scheme is much more general and requires quasi group-velocity
mathing between different optical pulses. This can be achieved in different
settings, and can, e.g., already been found in an early work of Hasegawa
\cite{Hasegawa:OL:1980}, where a strong incoherent XPM interaction between
different components of a multimode optical pulse has been considered.
At this point, we would also like to emphasize that these pulse compounds are
different from usual soliton molecules, which can be realized by dispersion
engineering in the framework of a standard NSE \cite{Stratmann:PRL:2005},
characterized by two pulses separated by a fixed temporal delay and stabilized
by a phase relation between both pulses \cite{Hause:PRA:2008}.
%
%


Here, we review the rich dynamical behavior of two-color pulse compounds, which
consist of two group-velocity matched subpulses in distinct domains of
anomalous dispersion, with frequency loci separated by a vast frequency gap. 
First, we will demonstrate paradigmatic propagation scenarios that demonstrate
photonic meta-atoms, arising in the limiting case where the pulse compounds
consist of an intense trapping pulse, given by a soliton, and a weak trapped
pulse.
Then, we will address the case where both subpulses have similar amplitudes, so
that their mutual XPM induced confining action results in the formation of a
narrow two-color soliton molecule.
Finally, we show that non-stationary dynamics of the subpulses results in the
emission of resonant radiation, and we show how the location of the newly
generated frequencies depends on the $z$-periodic amplitude and width
variations of the oscillating soliton molecule.

The article is organized as follows.
In Sec.~\ref{sec:model} we discuss the propagation model used for our
theoretical investigations of two-color meta-atoms and soliton molecules, and
detail the numerical methods employed for their simulation and analysis.
In Sec.~\ref{sec:atoms} we demonstrate the ability of solitons to act as
attractive potential wells that can host trapped states, and probe
the stability of the resulting photonic meta-atoms with respect to a
group-velocity mismatch between the trapping soliton and the trapped state.
In Sec.~\ref{sec:molecules} we derive a simplified model that yields
simultaneous solutions for the subpulses that make up a two-color soliton
molecule and show that theses solutions entail the two-color soliton pairs
derived in Ref.~\cite{Melchert:OL:2021}. 
We perturb these pulse compounds by increasing their initial amplitude, which
results in periodic amplitude and width oscillations, and triggers the
generation of resonant multi-frequency radiation with a complex stucture that
can be precisely predicted theoretically. 
Section~\ref{sec:conclusion} concludes with a summary.

\section{Model and methods}
\label{sec:model}

\paragraph{Propagation model}
In order to study the propagation dynamics of nonlinear photonic meta-atoms and
two-color soliton molecules, we consider a modified nonlinear Schrödinger
equation (NSE) of the form 
\begin{eqnarray}
i \partial_z A = \left(\frac{\beta_2}{2} \partial_t^2 - \frac{\beta_4}{24}\partial_t^4\right) A - \gamma |A|^2 A, \label{eq:NSE}
\end{eqnarray}
describing the single-mode propagation of a complex-valued field $A\equiv
A(z,t)$, on a periodic temporal domain of extent $T$ for the boundary condition
$A(z,-T/2)=A(z,T/2)$. 
The linear part of Eq.~(\ref{eq:NSE}) includes higher orders of dispersion,
with $\beta_2>0$ (in units of $\mathrm{fs^2/\mu m}$) a positive-valued
group-velocity dispersion coefficient, and $\beta_4<0$ ($\mathrm{fs^4/\mu m}$)
a negative-valued fourth-order dispersion coefficient. 
The nonlinear part of Eq.~(\ref{eq:NSE}) includes a positive-valued scalar
nonlinear coefficient $\gamma$ ($\mathrm{W^{-1}/\mu m}$).
%
%
%
%
Considering the discrete set of angular frequency detunings $\Omega\in
\frac{2\pi}{T}\mathbb{Z}$, the transform-pair 
\begin{subequations} \label{eq:FT}
\begin{align}
&A_\Omega(z)=\mathsf{F}[A(z,t)] \equiv \frac{1}{T} \int_{-T/2}^{T/2} A(z,t)\,e^{i\Omega t}~{\rm{d}}t, \label{eq:FT_FT}\\
&A(z,t) = \mathsf{F}^{-1}[A_\Omega(z)] \equiv \sum_{\Omega} A_\Omega(z)\,e^{-i\Omega t}, \label{eq:FT_IFT}
\end{align}
\end{subequations}
specifies a Fourier transform [Eq.~(\ref{eq:FT_FT})], and the corresponding
inverse [Eq.~(\ref{eq:FT_IFT})], relating the field envelope $A(z,t)$ to the
spectral envelope $A_\Omega(z)$.

\begin{figure}[t!]
\centerline{
\includegraphics[width=0.5\linewidth]{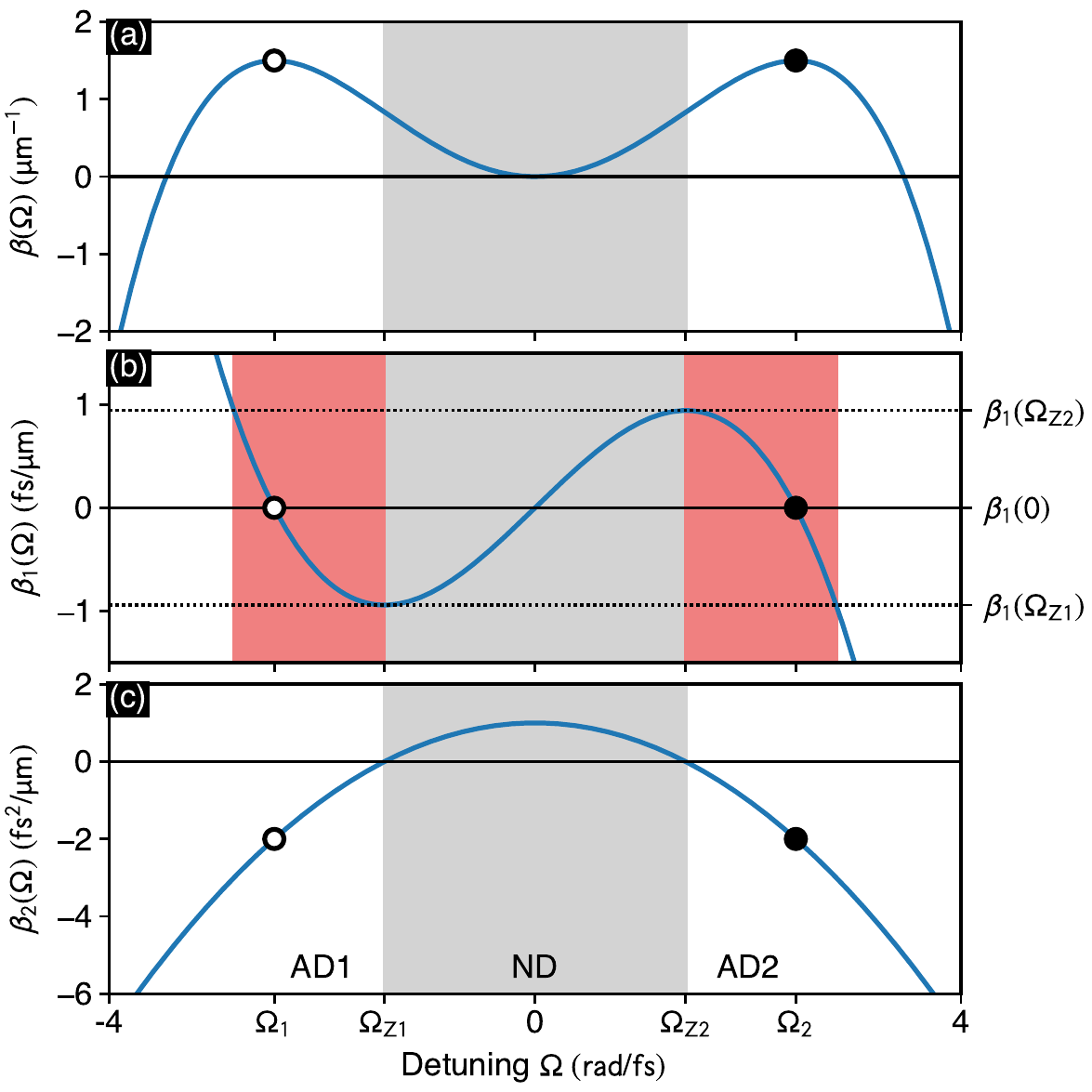}
}
\caption{Details of the frequency-dependent propagation constant supporting
nonlinear-photonic meta-atoms and two-color soliton molecules.
(a) Propagation constant,
(b) inverse group velocity, and,
(c) group-velocity dispersion.
In (c), AD1 and AD2 label two distinct domains of anomalous dispersion,
separated by an extended domain of normal dispersion (labeled ND). 
In (a-c), the domain of normal dispersion is shaded gray.
Zero-dispersion points are labeled $\Omega_{\rm{Z1}}$ and $\Omega_{\rm{Z2}}$.
In (b), the frequency range shaded in red allows for group-velocity matching 
of two modes with loci in AD1 and AD2.
Open circle (labeled $\Omega_1$) and filled circle (labeled $\Omega_2$)
indicate such a pair of group-velocity matched frequencies.
\label{fig:01}}
\end{figure}

\paragraph{Propagation constant}
Using the identity $\partial_t^n\,e^{-i \Omega t} = (-i\Omega)^n\,e^{-i\Omega
t}$ of the spectral derivative,\footnote{
Let us note that the ``-''-sign in the bracket on the right-hand-side of the
preceding identity reflects the sign-choice of the plane-wave basis in
Eqs.~(\ref{eq:FT}). 
This has to be taken into account when using scientific computing tools such
as, e.g., Python's scipy package \cite{Scipy,Virtanen:NM:2020}, where readily
available routines for spectral derivative exist that implement a different
sign-choice for the pair of Fourier-transforms.
}
%
%
the frequency-domain representation of the propagation
constant is given by the polynomial expression 
\begin{subequations}\label{eq:betas}
\begin{align}
\beta(\Omega) =
\frac{\beta_2}{2}\Omega^2 + \frac{\beta_4}{24} \Omega^4. \label{eq:beta} 
\end{align}
%
%
%
The frequency-dependent inverse group-velocity of a mode at detuning $\Omega$
reads 
\begin{align}
\beta_1(\Omega)\equiv \partial_\Omega \beta(\Omega)= \beta_2 \Omega + \frac{\beta_4}{6} \Omega^3 \label{eq:beta1},
\end{align}
with group-velocity (GV) $v_g(\Omega)=1/\beta_1(\Omega)$, and the
group-velocity dispersion (GVD) is given by 
\begin{align}
\beta_2(\Omega)\equiv \partial_\Omega^2 \beta(\Omega) = \beta_2 + \frac{\beta_4}{2}\Omega^2. \label{eq:beta2}
\end{align}
\end{subequations}
Subsequently, we use the parameter values $\beta_2 = 1~\mathrm{fs^2/\mu m}$,
and $\beta_4= -1~\mathrm{fs^4/\mu m}$, resulting in the model dispersion
characteristics shown in Fig.~\ref{fig:01}.
For the nonlinear coefficient in Eq.~(\ref{eq:NSE}) we use $\gamma =
1~\mathrm{W^{-1}/\mu m}$.
%
%
%
As evident from Fig.~\ref{fig:01}(c), the GVD profile
Eq.~(\ref{eq:beta2}) has a concave downward shape with two zero-dispersion
points, defined by the condition $\beta_2(\Omega)\stackrel{!}{=}0$, located at
$\Omega_{\rm{Z1},\rm{Z2}}=\mp \sqrt{2\beta_2/|\beta_4|}=\mp
\sqrt{2}\,\mathrm{rad/fs}$.
It exhibits anomalous dispersion for $\Omega<\Omega_{\rm{Z1}}$ as well as 
for $\Omega > \Omega_{\rm{Z2}}$.
The interjacent frequency range $\Omega_{\rm{Z1}}< \Omega < \Omega_{\rm{Z2}}$
exhibits normal dispersion.
Inspecting the inverse group velocity shown in Fig.~\ref{fig:01}(b), it can be
seen that two frequencies are GV matched to $\Omega = 0$.  Due to the symmetry
of the propagation constant, these are given by the pair $\Omega_1 =
-\Omega_2=-\sqrt{6 \beta_2/|\beta_4|}\approx -2.828~\mathrm{rad/fs}$, uniquely
characterized by $\beta(\Omega_1)=\beta(\Omega_2)$ and indicated by the open
and filled circles in Fig.~\ref{fig:01}.
In fact, for the considered propagation constant, GV matching of three distinct
modes can be realized as long as the frequency loci in AD1 and AD2 lie within
the range of frequencies shaded in red in Fig.~\ref{fig:01}(b).
%
Let us note that while the type of GV matching for two optical pulses at vastly
different center frequencies, supported by the propagation constant
Eq.~(\ref{eq:beta}), it is methodologically different from the type of GV
matching that supports quasi co-propagation of different modes with similar
frequencies \cite{Hasegawa:OL:1980}.
%
%
Nevertheless, both allow for quasi co-propagation of optical pulses under
different circumstances, supporting similar XPM induced propagation effects.
In our case, quasi group-velocity matched propagation of optical pulses across
a vast frequency gap is possible, enabled by a tailored propagation constant
with multiple zero-dispersion points.
%
Further, the considered mechanism of GV matching differs from that in
Ref.~\cite{Saleh:PRA:2013}, wherein two pulses at the same central frequency
but different polarization states were assumed to be launched in the anomalous
dispersion regime of a hollow-core photonic crystal fiber filled with a noble
gas.
%
%
The mathematical structure of Eq.~(\ref{eq:NSE}) and the above choice of
parameters yields a very basic setting supporting the stable propagation of 
nonlinear photonic meta-atoms and two-color soliton molecules.
In fact, the two-parameter GVD curve shown in Fig.~\ref{fig:01}(c) is a
simplified model of the dispersion considered earlier in
Ref.~\cite{Melchert:PRL:2019}, wherein two-color soliton molecules were first
demonstrated, and is similar to the setting considered in
Ref.~\cite{Tam:PRA:2020}, wherein generalized dispersion Kerr solitons were
described comprehensively.
However, let us note that the phenomena reported below are not limited to the
particular choice of the above parameters and persist even in the presence of
perturbations such as pulse self-steepening
\cite{Melchert:SR:2021,Melchert:OL:2021}, which can be accounted for by
replacing $\gamma \rightarrow \gamma(\Omega)$ in the nonlinear part of
Eq.~(\ref{eq:NSE}), and -- with some reservation -- a self-frequency shift
caused by the Raman effect \cite{Willms:PRA:2022}.

%
\paragraph{Propagation algorithm}
For our pulse propagation simulations in terms of Eq.~(\ref{eq:NSE}), we employ
the ``Conservation quantity error'' method (CQE)
\cite{Heidt:JLT:2009,Melchert:CPC:2022}. 
It maintains an adaptive $z$-propagation stepsize $h$, and uses a conservation
law of the underlying propagation equation to guide stepsize selection.
Specifically, we here use the relative error 
\begin{eqnarray}
\delta_{E}(z) =  \frac{|E(z+h) - E(z)|}{E(z)}, \label{eq:dE}
\end{eqnarray} 
where $E$ is the total energy, conserved by Eq.~(\ref{eq:NSE}).
Employing Parseval's identity for Eqs.~(\ref{eq:FT})
\cite{NR:BOOK:2007,Weideman:SIAM:1986}, the total energy in the time and
frequency domains is given by
\begin{eqnarray}
E(z) = \int_{-T/2}^{T/2} |A(z,t)|^2~{\rm{d}}t = T \sum_\Omega |A_\Omega(z)|^2, \label{eq:E}
\end{eqnarray}
with instantaneous power $|A(z,t)|^2$ ($\mathrm{W=J/s}$), and power spectrum
$|A_\Omega(z)|^2$ ($\mathrm{W}$).
The CQE method is designed to keep the relative error $\delta_{E}$ within the
goal error range $(0.1\,\delta_{\rm{G}}, \delta_{\rm{G}})$, for a preset local
goal error $\delta_{\rm{G}}$ (throughout our numerical experiments we set
$\delta_{\rm{G}}=10^{-10}$). This is accomplished by decreasing the stepsize
$h$ when necessary while increasing $h$ when possible. 
To advance the field from position $z$ to $z+h$, the CQE uses the
``Fourth-order Runge-Kutta in the interaction picture'' (RK4IP) method
\cite{Hult:JLT:2007}.
The ability of the algorithm to increase or decrease the stepsize is most
valuable when the propagation of an initial condition results in a rapid change
of the pulse intensities over short propagation distances.
%
Nevertheless, if one is willing to accept an increased running time resulting
from an integration scheme with fixed stepsize, usual split-step Fourier
methods \cite{Taha:JCP:1984,Weideman:SIAM:1986,DeVries:AIP:1987} will work
similarly well.

\paragraph{Spectrograms}
To assess the time-frequency interrelations within the field $A(z,t)$ at a
selected propagation distance $z$, we use the spectrogram
\cite{Melchert:SFX:2019,Cohen:IEEE:1989,Dudley:OL:2002b}
\begin{equation}
P_{S}(t,\Omega;z) = \frac{1}{2 \pi} \left|\int_{-T/2}^{T/2} A(z,t^\prime)h(t^\prime-t) e^{-i \Omega t^\prime}~{\rm d}t^\prime\right|^2. \label{eq:PS}
\end{equation}
To localize the field in time, we use a hyperbolic-secant window function
$h(x)={\mathrm{sech}}(x/\sigma)$ with width parameter $\sigma$.
%

\paragraph{Incoherently coupled pulse pairs}
%
%
To facilitate a simplified description of two-color pulse compounds in the form 
\begin{align}
A(z,t) = A_1(z,t) \, e^{-i \Omega_1 t} + A_2(z,t) \, e^{-i \Omega_2 t}, \label{eq:A1A2ansatz}
\end{align}
in which two quasi group-velocity-matched subpulses $A_1\equiv A_1(z,t)$ and
$A_2\equiv A_2(z,t)$ exist at the frequency gap
$\Omega_{\rm{gap}} = |\Omega_2-\Omega_1|$, it is convenient to consider the two
coupled nonlinear Schrödinger equations (CNSEs)
\cite{Agrawal:BOOK:2019,Tan:CSF:2000,Tan:JAS:1995} 
\begin{subequations}\label{eq:CNSE}
\begin{align}
&i \partial_z \,A_1 + \beta_0^{\prime}\, A_1 - i \beta_1^{\prime} \partial_t\,A_1 - \frac{\beta_2^\prime}{2} \partial_t^2 \,A_1 + \gamma^\prime \left( |A_1|^2 + 2 |A_2|^2 \right) A_1=0, \label{eq:CNSE1} \\ 
&i \partial_z \,A_2 + \beta_0^{\prime\prime}\, A_2- i \beta_1^{\prime\prime} \partial_t\,A_2 - \frac{\beta_2^{\prime\prime}}{2} \partial_t^2 \,A_2 + \gamma^{\prime\prime} \left( |A_2|^2 + 2 |A_1|^2 \right) A_2=0.
\label{eq:CNSE2}
\end{align}
\end{subequations}
%
%
The parameters in Eqs.~(\ref{eq:CNSE}) are related to Eqs.~(\ref{eq:betas})
through
$\beta_0^{\prime}=\beta(\Omega_1)$, 
$\beta_0^{\prime\prime}=\beta(\Omega_2)$,
$\beta_1^{\prime}=\beta_1(\Omega_1)$,
$\beta_1^{\prime\prime}=\beta_1(\Omega_2)$,
$\beta_2^{\prime} = \beta_2(\Omega_1)$, 
$\beta_2^{\prime\prime} = \beta_2(\Omega_2)$, and,
$\gamma^{\prime}=\gamma^{\prime\prime}=\gamma$.
The mismatch of inverse GV for both subpulses is given by $\Delta \beta_1
\equiv |\beta_1^{\prime\prime}- \beta_1^{\prime}|$.  For specific choices of
the detunings $\Omega_1$ and $\Omega_2$, exact GV matching, signaled by
$\Delta \beta_1 = 0$, can be achieved.
%
%
In contrast to Eq.~(\ref{eq:NSE}), the incoherently coupled
Eqs.~(\ref{eq:CNSE}) neglect higher-orders of dispersion within their linear
parts, as well as rapidly varying four-wave-mixing terms within their nonlinear
parts.
%
%
The mutual interaction of both subpulses is taken into account via XPM.
As evident from Eq.~(\ref{eq:CNSE1}), pulse $A_1$ can be viewed as being
exposed to a total potential field of the form $V_1 \equiv \gamma^{\prime}
(|A_1|^2 + 2 |A_2|^2)$, entailing the effects of SPM and XPM.  Likewise, $A_2$
is exposed to the potential field $V_2 \equiv \gamma^{\prime\prime} (|A_2|^2 +
2 |A_1|^2)$.
As we will show in Sects.~(\ref{sec:atoms}), (\ref{sec:molecules}), the
potential fields $V_1$ and $V_2$ yield attractive potentials that enable the
mutual trapping of both subpulses.
%
%
Subsequently we take $\Omega_1$ and $\Omega_2$ as indicated in
Fig.~\ref{fig:01}, so that the above parameters are given by
$\beta_0^{\prime}=\beta_0^{\prime\prime}=1.33~\mathrm{\mu m^{-1}}$,
$\beta_1^{\prime}=\beta_1^{\prime\prime}=0$, $\beta_2^{\prime} =
\beta_2^{\prime\prime}= -2~\mathrm{fs^2/\mu m}$, and, 
$\gamma^{\prime}=\gamma^{\prime\prime}=1~\mathrm{W^{-1}/\mu m}$. 
For a more general description of simultaneous solutions in the form of
Eq.~(\ref{eq:A1A2ansatz}), we will continue to refer to the nonlinear
coefficients in Eqs.~(\ref{eq:CNSE}) as $\gamma^{\prime}$
[Eq.~(\ref{eq:CNSE1})] and $\gamma^{\prime\prime}$ [Eq.~(\ref{eq:CNSE2})].
In addition, the scalar factors
$\beta_0^{\prime}=\beta_0^{\prime\prime}\equiv\beta_0$ can be removed by a
common linear transformation $A_{1,2} \rightarrow A_{1,2} e^{i\beta_0 z}$,
which does not affect the $z$-propagation dynamics of the interacting pulses.

%
%
Let us note that, in general, higher-orders of dispersion within a modified NSE
can cause a solitary wave to shed resonant radiation \cite{Akhmediev:PRA:1995},
and can result in a modification of its group-velocity
\cite{Akhmediev:PRA:1995,Pickartz:PRA:2016}.
These types of perturbations are neglected by Eqs.~(\ref{eq:CNSE}), which can
be justified in the limit where the subpulse separation $\Omega_{\rm{gap}}$ is
large and their spectra are sufficiently narrow. 
Moreover, in case of a frequency dependent coefficient function
$\gamma(\Omega)$, $\gamma^{\prime}=\gamma(\Omega_1)$ and
$\gamma^{\prime\prime}=\gamma(\Omega_2)$ in Eqs.~(\ref{eq:CNSE}).
Let us point out that, in the presence of a linear variation of $\gamma$, a
solitary wave exhibits a further modification of its group-velocity
\cite{Haus:OL:2001}, an effect neglected by Eqs.~(\ref{eq:CNSE}).
It is important to bear these perturbation effects in mind when comparing 
results based on Eqs.~(\ref{eq:CNSE}) to numerical simulations in terms 
of the full model Eq.~(\ref{eq:NSE}).

%
%
We can relate the above trapping mechanism for two-color pulse compounds to the
mechanism enabling the self-confinement of a multimode optical pulses in a
multimode fiber, discussed by Hasegawa as early as 1980
\cite{Hasegawa:OL:1980}.
Therein, Hasegawa considered a propagation equation of the nonlinear
Schrödinger type for a multimodal pulse, where the nonlinear change of the
refractive index, felt by an individual mode, depends on the total intensity of
the multimodal pulse.  
This results in coupled equations for the different modes, wherein an
individual mode perceives the intensity of the total pulse as a potential
field.
If the considered mode is subject to anomalous dispersion, the potential is
attractive. 
Based on the expectation that if the velocity mismatch between a given mode and
the potential is smaller than the escape velocity, the potential has the
ability to trap the mode, he derived a condition for self-confinement of the
multimode pulse.
While the results in Ref.~\cite{Hasegawa:OL:1980} are valid for multimodal
optical pulses composed of possibly many modes, the simplified modeling
approach given by Eqs.~(\ref{eq:CNSE}) considers only two subpulses.
Meanwhile, an extension of the above approach to pulse compounds with three and
more subpulses has been accomplished
\cite{Willms:PRA:2022,Lourdesamy:JOSA:2023}.
%

Given the ansatz for two-color pulse compounds in the form of
Eq.~(\ref{eq:A1A2ansatz}), initial conditions $A_0(t)\equiv A(z=0,t)$ that
specify nonlinear photonic meta-atoms and two-color soliton molecules in terms
of the subpulses $A_1$ and $A_2$ are different in some respects and are
discussed separately in Sect.~\ref{sec:atoms}, and Sect.~\ref{sec:molecules}.
Subsequently, we demonstrate the self-consistent $z$-propagation dynamics of
these pulse compound, originally reported in
Refs.~\cite{Melchert:PRL:2019,Tam:PRA:2020,Melchert:OL:2021,Melchert:OL:2023,Melchert:NJP:2023},
as well as their breakup in response to sufficiently large GV mismatches
between both subpulses, originally reported in Ref.~\cite{Melchert:SR:2021}, in
terms of numerical simulations governed by the full model Eq.~(\ref{eq:NSE}).
These numerical results demonstrate several theoretical findings reported by
Hasegawa \cite{Hasegawa:OL:1980}, applied to the concept of two-color pulse
compounds.

In passing, let us stress that coupled equations of the form of
Eqs.~(\ref{eq:CNSE}) comprise a much-used theoretical instrument for studying
mutually bound solitons
\cite{Ueda:PRA:1990,Menyuk:JOSAB:1988,Afanasjev:PZ:1988,Trillo:OL:88,Afanasjev:JQE:1989,Afanasjev:OL:1989,Mesentsev:OL:1992,Akhmediev:Chaos:2000}.

\section{Nonlinear-photonics meta-atoms}
\label{sec:atoms}

\paragraph{Description of stationary trapped states}
Subsequently we look for stationary solutions in the form of
Eq.~(\ref{eq:A1A2ansatz}) under the additional constraint ${\rm{max}}(|A_2|)\ll
{\rm{max}}(|A_1|)$.
This allow to decouple Eqs.~(\ref{eq:CNSE}) and enables direct optical
analogues of quantum mechanical bound-states
\cite{Landau:BOOK:1981,Melchert:PRL:2019,Melchert:OL:2023}.
Therefore, we assume the resulting two-color pulse compounds to consist of a
strong trapping pulse, given by a solitary wave (S) at detuning
$\Omega_{\rm{S}}\equiv\Omega_1$, and a weak trapped pulse (TR) at detuning
$\Omega_{\rm{TR}}\equiv\Omega_2$.
For the solitary wave part of the total pulse we neglect the XPM contribution
in the nonlinear part of Eq.~(\ref{eq:CNSE1}) and assume 
\begin{align}
A_1(z,t) = U_{\rm{S}}(t) \, e^{i \kappa^{\prime} z},\quad\text{with}\quad U_{\rm{S}}(t)=\sqrt{P_0}\,{\rm{sech}}\left(\frac{t}{t_0}\right),\label{eq:Sansatz}
\end{align}
wherein
$P_0=|\beta_2^{\prime}|/(\gamma^\prime t_0^2)$, and $\kappa^{\prime} = \beta_0 +
\gamma^{\prime} P_0/2$.
Neglecting the SPM contribution in the nonlinear part of Eq.~(\ref{eq:CNSE2})
and making the ansatz 
\begin{align}
A_2(z,t) = \phi(t) \, e^{i \kappa^{\prime\prime} z},\label{eq:TRansatz}
\end{align}
the envelope $\phi(t)$ of a weak stationary trapped state is determined by the
Schrödinger type eigenvalue problem
\begin{align}
\left(-\frac{|\beta_2^{\prime \prime}|}{2} \frac{{\rm{d}}^2}{{\rm{d}}t^2} + V_S(t) 
\right) \, \phi_n(t) = \kappa_n \, \phi_n(t). \label{eq:EVproblem}
\end{align}
Therein, the solitary wave enters as a  stationary attractive potential well
$V_{\rm{S}}(t) = - 2 \gamma^{\prime\prime} P_0 \, {\rm{sech}}^2(t/t_0)$.
Hence, as pointed out above and discussed in the context of multimode optical
pulses in glass fibers in Ref.~\cite{Hasegawa:OL:1980}, a weak pulse can be
attracted by the intensity of the entire pulse if it exists in a domain of
anomalous dispersion. Due to $\beta_2^{\prime \prime}<0$, this condition is met
in the considered case.
%
%
%
In analogy to the  \mbox{${\mathrm{sech}}^2$-potential} in one-dimensional
quantum scattering theory we may equivalently write the solitary-wave induced
potential as \cite{Landau:BOOK:1981}
\begin{align}
V_{\rm{S}}(t)=-\nu\,(\nu+1)\,\frac{|\beta_2^{\prime\prime}|}{2t_0^{2}}\,{\mathrm{sech}}^2\left(\frac{t}{t_0}\right),
\quad \text{with} \quad 
\nu=-\frac{1}{2} + \left(\frac{1}{4} + 4 \left|
\frac{\gamma^{\prime\prime}}{\gamma^{\prime}}\frac{\beta_2^{\prime}}{\beta_2^{\prime\prime}}\right|
\right)^{1/2}. \label{eq:VPT}
\end{align}
%
Moreover, due to the particular shape of the trapping potential, the eigenvalue
problem Eq.~(\ref{eq:EVproblem}) can even be solved exactly
\cite{Landau:BOOK:1981,Lekner:AJP:2007}.
The number of trapped states of the potential in Eq.~(\ref{eq:VPT}) is given by
$N_{\rm{TR}}=\lfloor \nu \rfloor + 1$, where $\lfloor \nu \rfloor$ is the
integer part of the strength-parameter $\nu$.
From the analogy to the quantum mechanical scattering problem
\cite{Lekner:AJP:2007}, the real-valued wavenumber eigenvalues can directly be
stated as 
\begin{align}
\kappa_n=-\frac{|\beta_2^{\prime\prime}|} {2t_0^2}\,(\nu-n)^2, \quad
\text{for} \quad n=0,\ldots,\lfloor \nu \rfloor. \label{eq:kappan}
\end{align}
For a given value of $n$, they are related to Eq.~(\ref{eq:TRansatz}) through
$\kappa^{\prime\prime} = \beta_0 - \kappa_n$.
To each eigenvalue corresponds an eigenfunction $\phi_n$ with $n$ zeros,
specifying the $(n+1)$-th fundamental solution of the eigenvalue problem
Eq.~(\ref{eq:EVproblem}).  
These solutions constitute the weak trapped states of the potential
$V_{\rm{S}}$.
Referring to the Gaussian hypergeometric function as ${_2F_1}$
\cite{Abramowitz:BOOK:1972}, and abbreviating $a_n=\tfrac{1}{2}(1+n)$ and
$b_n=\tfrac{1}{2}(2\nu+1-n)$, they can be stated in closed form as
\cite{Lekner:AJP:2007} 
\begin{align}
\phi_{n}(t) &= 
\begin{cases}
\cosh^{\nu+1}\left(\frac{t}{t_0}\right)~{_2F_1}\left[ a_n, b_n; \frac{1}{2}; -\sinh^2\left(\frac{t}{t_0}\right) \right],
& \text{for even $n$},\\
\cosh^{\nu+1}\left(\frac{t}{t_0}\right)~\sinh\left(\frac{t}{t_0}\right)~{_2F_1}\left[ a_n+\frac{1}{2}, b_n+\frac{1}{2}; \frac{3}{2}; -\sinh^2\left(\frac{t}{t_0}\right) \right],
& \text{for odd $n$}.\\
\end{cases}\label{eq:eigFunc}
\end{align}
Let us note that, as evident from the potential strength parameter $\nu$ in
Eq.~(\ref{eq:VPT}), the number $N_{\rm{TR}}$ of trapped states is uniquely
defined by the four parameters $\beta_2^{\prime}$, $\beta_2^{\prime\prime}$,
$\gamma^{\prime}$, and $\gamma^{\prime \prime}$. It is not affected by the
duration $t_0$ of the trapping potential, which, according to
Eq.~(\ref{eq:kappan}), codetermines the value of the wavenumber eigenvalue of a
fundamental solution.

\paragraph{Analogy to quantum mechanics}
The eigenvalue problem Eq.~(\ref{eq:EVproblem}) suggests an analogy to quantum
mechanics, wherein a fundamental solution $\phi_n$ represents the wavefunction
of a fictitious particle of mass $m=|\beta_2^{\prime\prime}|^{-1}$, confined to
a localized, ${\rm{sech}}^2$-shaped trapping potential $V_{\rm{S}}$.  
The discrete variable $n=0,\ldots,\lfloor \nu \rfloor$ resembles a principal
quantum number that labels solutions with distinct wavenumbers, and the number
of trapped state $N_{\rm{TR}}$ is similar to an atomic number. 
Consequently, a bare soliton, with none of its trapped states occupied,
resembles the nucleus of an one-dimensional atom.  By this analogy, a soliton
along with its trapped states represents a nonlinear-photonics meta-atom.

\begin{figure}[t!]
\includegraphics[width=\linewidth]{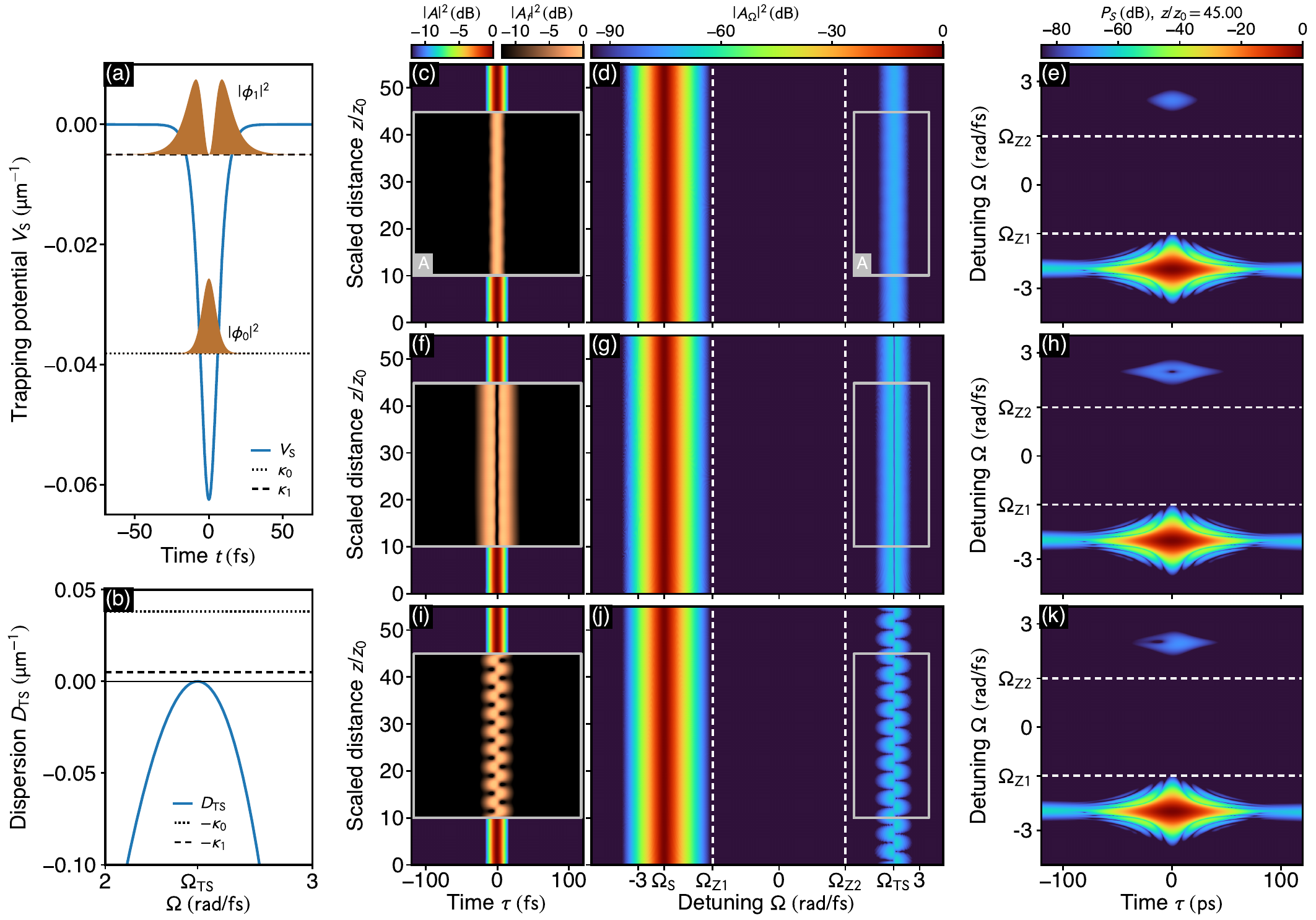}
\caption{Solitary-wave induced potential well exhibiting two trapped states.
(a) Trapping potential $V_{\rm{S}}$, wavenumber eigenvalues $\kappa_n$, and 
squared magnitude $|\phi_n|^2$ of trapped state eigenfunctions for $n=0,1$.
(b) Dispersion profile $D_{\rm{TS}}(\Omega)$ in the vicinity of the trapped
state center frequency $\Omega_{\rm{TS}}$,  
(c) Time-domain propagation dynamics of the soliton and its lowest lying
trapped state for $n=0$. The propagation distance is scaled by the soliton
period $z_0 = (\pi/2) (t_0^2/{\beta_2^{\prime}})\approx 50~\mathrm{\mu m}$. 
(d) Corresponding spectrum. The inverse Fourier transform of the part of the
spectrum enclosed by the box (labeled A) in (d) is shown in the box (labeled A)
in (c), providing a filtered view of the trapped state while leaving out the
soliton part of the total pulse.
(e) Spectrogram of the total pulse at $z/z_0=45$ for $\sigma=8~\mathrm{fs}$. 
(f,g,h) Same as (c,d,e) for the trapped state with $n=1$.
(i,j,k) Same as (c,d,e) for a superposition of both trapped states.
Movies of the propagation dynamics are provided as supplementary material 
under Ref.~\cite{SuppMat:Zenodo:2023}.
\label{fig:02}}
\end{figure}

%

\subsection{Stable propagation of trapped states}
\label{sec:trapped_states}

Subsequently, we discuss the propagation dynamics of a nonlinear-photonics 
meta-atom with the ability to host two trapped states.
More precisely, we consider an example for 
$\Omega_{\rm{S}}=-2.828\,\mathrm{rad/fs}$ and $t_0=8\,\mathrm{fs}$, with
$\Omega_{\rm{TR}}=2.828\,\mathrm{rad/fs}$ and $\nu \approx 1.566$.
The resulting trapping potential and both its trapped states are shown in
Fig.~\ref{fig:02}(a).
In this case, the wavenumber eigenvalues are $(\kappa_0,\,\kappa_1)=(-0.0382,
-0.0050)\,\mathrm{\mu m^{-1}}$, and the corresponding fundamental solutions
take the simple form 
\begin{subequations}\label{eq:phi01}
\begin{align}
\phi_0(t) &= {\mathrm{sech}}^\nu\left(\frac{t}{t_0}\right), \quad \text{and}, \label{eq:phi0}\\
\phi_1(t) &= {\mathrm{sech}}^{\nu-1}\left(\frac{t}{t_0}\right)\,{\mathrm{tanh}}\left(\frac{t}{t_0}\right). \label{eq:phi1}
\end{align}
\end{subequations}
As evident in Fig.~\ref{fig:02}(b), in the vicinity of $\Omega_{\mathrm{TR}}$
and due to $\kappa^{\prime\prime}>0$ [Eq.~(\ref{eq:TRansatz})], a finite
wavenumber-gap separates each trapped state from linear waves bound to the
dispersion curve $D_{\rm{TR}}(\Omega) \equiv \beta(\Omega) -
\beta(\Omega_{\rm{TR}}) - \beta_1(\Omega_{\rm{TR}}) (\Omega-\Omega_{\rm{TR}})
<0 $. Therefore, we expect that trapped states composed by
Eqs.~(\ref{eq:phi01}) propagate in a stable manner.
For the lowest lying trapped state, having order $n=0$, this is demonstrated in
Figs.~\ref{fig:02}(c,d).
These figures summarize pulse propagation simulations in terms of the
modified NSE~(\ref{eq:NSE}), using an initial condition of the form of
Eq.~(\ref{eq:A1A2ansatz}) with $A_1$ as in Eq.~(\ref{eq:Sansatz}), and $A_2$ as
in Eq.~(\ref{eq:TRansatz}) with $\phi(t)=\sqrt{10^{-7}P_0}\,\phi_0(t)$.
In the time-domain propagation dynamics, shown in Fig.~\ref{fig:02}(c), a small
drift of the soliton, caused by higher orders of dispersion at
$\Omega_{\rm{S}}$ [see Fig.~\ref{fig:01}], is accounted for by shifting to a
moving frame of reference with time coordinate $\tau = t - \tilde{\beta}_{1} z$
and $\tilde{\beta}_{1} = 0.00637~\mathrm{fs/\mu m}$.
In Fig.~\ref{fig:02}(d), the vast frequency gap between the soliton and 
the trapped state is clearly visible. 
By means of an inverse Fourier transform of the frequency components belonging
to the trapped state [box labeled A in Fig.~\ref{fig:02}(d)], an unhindered
``filtered view'' of the  time-domain propagation dynamics of the trapped state
is possible  [box labeled A in Fig.~\ref{fig:02}(c)].
A spectrogram, providing a time-frequency view of the field at $z/z_0=45$, is
shown in Fig.~\ref{fig:02}(d). 
The stable propagation of a trapped state with $n=1$ for
$\phi(t)=\sqrt{10^{-7}P_0}\,\phi_1(t)$, is detailed in Figs.~\ref{fig:02}(f-h).
Finally, the simultaneous propagation of a superposition of both trapped states
in the form $\phi(t) = \sqrt{10^{-7}P_0}\,[ \phi_0(t) + 5 \phi_1(t)]$ is shown
in Figs.~\ref{fig:02}(i-k).
The $z$-periodicity of the beating pattern visible in the time-domain
propagation dynamics in Fig.~\ref{fig:02}(i), is a result of the different
wavenumber eigenvalues of the trapped states, and is determined by
$z_{\rm{p}}=2\pi/|\kappa_1-\kappa_0|\approx 189~\mathrm{\mu m}$
($z_{\rm{p}}/z_0 \approx 3.8$).
Thus, the coherent superposition of trapped states exhibits Rabi-type
oscillations, similar to bound state dependent revival times in the quantum
recurrence of wave packets \cite{Bocchieri:PR:1957,Styer:AJP:2001}.

Let us note that, bearing in mind that the number of bound states $N_{\rm{TR}}$
is determined by the potential strength parameter $\nu$ in Eq.~(\ref{eq:VPT}),
a setup with a different number of bound states can be obtained as well.  This
is possible by fixing $\Omega_{\rm{S}}$ at some other feasible value, resulting
in a different group-velocity matched detuning $\Omega_{\rm{TR}}$, implying
different values of the parameters $\beta_2^{\prime}$,
$\beta_2^{\prime\prime}$, $\gamma^{\prime}$, and $\gamma^{\prime \prime}$.  
For example, keeping $t_0=8\,\mathrm{fs}$ but choosing
$\Omega_{\rm{S}}=-2.75~\mathrm{rad/fs}$ yields $\nu\approx 3.1$, resulting in a
potential well with the ability to host $N_{\rm{TR}}=4$ trapped states.  
In such a case, however, phase-matched transfer of energy from the trapped
states to dispersive waves within the domain of normal dispersion can be
efficient \cite{Babushkin:preprint:2022}.

%

\subsection{Trapping-to-escape transition caused by a group-velocity mismatch}

\begin{figure}[t!]
\includegraphics[width=\linewidth]{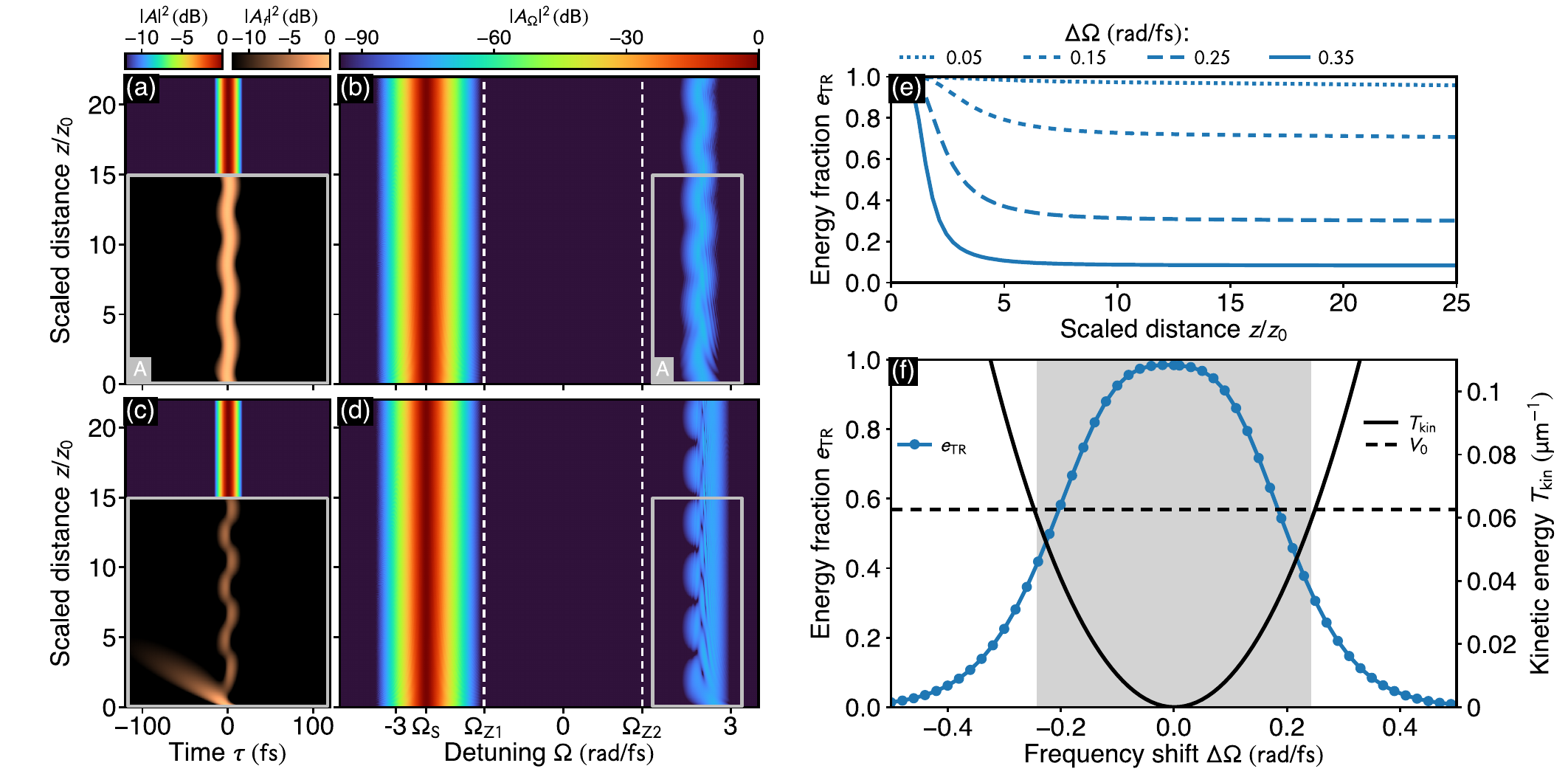}
\caption{Characterization of the transition from trapping to escape.
(a) Time-domain propagation dynamics of the soliton and its lowest lying
trapped state ($n=0$) shifted from $\Omega_{\rm{TS}}$ to
$\Omega_{\rm{TS}}+\Delta \Omega$ for $\Delta \Omega=0.05~\mathrm{(rad/fs)}$. 
(b) Corresponding spectrum. The inverse Fourier transform of the part of the
spectrum enclosed by the box (labeled A) in (b) is shown in the box (labeled A)
in (a). This provides a filtered view of the trapped state with the benefit of
leaving out the soliton part of the total pulse.
(c,d) Same as (a,b) for $\Delta \Omega=0.25~\mathrm{(rad/fs)}$.
(e) Fraction of trapped energy
as function of the propagation distance.
(f) Fraction of trapped energy as function of the trapped state center
frequency shift. 
Secondary ordinate shows the potential depth ($V_0$) as well as the kinetic 
energy $T_{\rm{kin}}$ of the fictitious classical particle.
Parameter range in which the particle cannot escape the well is shaded gray.
Movies of the propagation dynamics shown in (a-d) are provided as
supplementary material under Ref.~\cite{SuppMat:Zenodo:2023}.
\label{fig:03}}
\end{figure}

In the context of multimodal pulses in glass fibers in
Ref.~\cite{Hasegawa:OL:1980}, the attraction of a wave packet by a potential
well, created by the total pulse, was illustrated in terms of the kinetic
equations of a fictitious particle associated with the wave packet.  
From a classical mechanics point of view, in order to ensure trapping of the
wave packet by the total pulse, the velocity mismatch between the particle and
the potential needs to be smaller than the escape velocity of the potential.
Based on this view, and for a given velocity mismatch, the critical value of
the total pulse intensity, required to achieve self-confinement, was determined
\cite{Hasegawa:OL:1980}.
In the presented work, pulse propagation simulations, such as those reported in
Fig.~\ref{fig:02}, comprise a complementary approach to study the considered
XPM induced attraction effect.
Specifically, by keeping the detuning of the soliton fixed at
$\Omega_{\rm{S}}=\Omega_1$, but shifting the detuning of the trapped pulse to
$\Omega_{\rm{TR}} = \Omega_2 + \Delta \Omega$, we can enforce a group-velocity
mismatch between both pulses and probe the stability of the meta-atom.
For $\Delta \Omega > 0$ it is $\beta_1(\Omega_{\rm{S}}) >
\beta_1(\Omega_{\rm{TR}})$, see Fig.~\ref{fig:01}(b). Thus, in a reference
frame in which the soliton is stationary, the trapped state will initially have
the propensity to move towards smaller times.
This is demonstrated in Figs.~\ref{fig:03}(a,b) for the center frequency shift
$\Delta \Omega = 0.05~\mathrm{rad/fs}$.
To assess the fraction of energy of the trapped state that is retained within
the soliton induced potential well, we consider the quantity 
\begin{align}
e_{\rm{TR}}(z)\equiv \frac{E_{\rm{TR}}(z)}{E_{\rm{TR}}(0)},\quad \text{with} \quad E_{\rm{TR}}(z)= \int_{-10\,t_0}^{10\,t_0} |\phi(z,\tau)|^2~{\rm{d}}\tau. \label{eq:eTR}
\end{align}
As evident from Fig.~\ref{fig:03}(e), at $\Delta \Omega=0.05~\mathrm{rad/fs}$,
the trapped state is kept almost entirely within the well, i.e.\
$e_{\rm{TR}}\approx 1$.
In contrast, at $\Delta \Omega= 0.25~\mathrm{rad/fs}$, a major share of the
trapped pulse escapes the well during the initial propagation stage
[Figs.~\ref{fig:01}(c,d)], indicated by the small value $e_{\rm{TR}}\approx
0.3$ [Fig.~\ref{fig:03}(e)].
Let us note that, when viewing the considered pulse compounds as meta-atoms,
the quantity $1-e_{\rm{TR}}(z)$ specifies the fraction of trapped energy that
is radiated away, resembling an ionization probability for quantum mechanical
atoms.
A parameter study, detailing the dependence of $e_{\rm{TR}}$ as function of the
center frequency shift $\Delta \Omega$ is summarized in Fig.~\ref{fig:03}(f).
The transition from trapping to escape can be supplemented by an entirely
classical picture similar as in Ref.~\cite{Hasegawa:OL:1980}: from a classical
point of view we might expect that a particle, initially located at the center
of the well, remains confined to the well if its ``classical'' kinetic energy
$T_{\rm{kin}} = \tfrac{1}{2} m \Delta \beta_1^2
=\tfrac{1}{2}|\beta_2^{\prime\prime}|^{-1}  \left[\beta_1(\Omega_{\rm{S}}) -
\beta_1(\Omega_{\rm{TR}})\right]^2 $ does not exceed the well depth
$V_0=2\gamma^{\prime\prime} P_0$.
As evident from Fig.~\ref{fig:03}, the findings based on this classical picture
complement the results obtained in terms of direct simulations of the modified
NSE~(\ref{eq:NSE}) very well.
The above results clearly demonstrate the limits of stability of nonlinear
photonics meta-atoms with respect to a group-velocity mismatch between the
trapping soliton and the trapped state. These findings are consistent with our
previous results on the break-up dynamics of two-color pulse compounds
\cite{Melchert:SR:2021}.

\begin{figure}[t!]
\includegraphics[width=\linewidth]{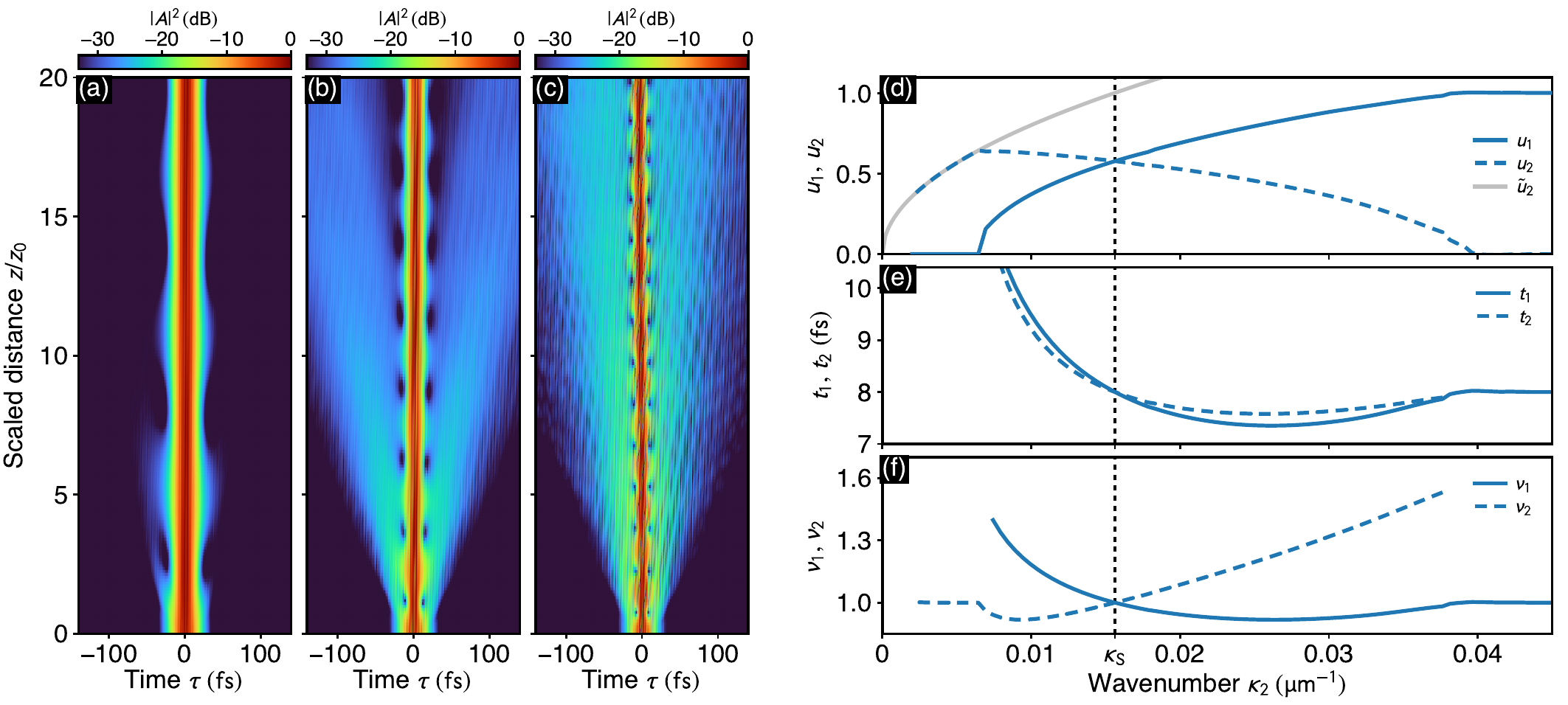}
\caption{Transition from trapping to tightly bound, molecule-like two-color
pulse compounds.
(a-c) Time-domain propagation dynamics arising from an initial condition of the
form $\phi(t)=r \sqrt{P_0}\,\phi_0(t)$ (see text).
(a) 
Trapped state for amplitude parameter $r=0.3$,
(b) $r=0.7$, and,
(c) $r=1$.
(d-f) Solutions of the coupled ODEs~(\ref{eq:ODEs}), fitted to functions of the
form $U_m=U_{0,m} {\rm{sech}}^{\nu_m}(t/t_m)$, for $m=1,2$.
(d) Scaled pulse amplitudes $u_n=U_{0,m}/\sqrt{P_0}$,
(e) pulse durations $t_n$, and,
(f) pulse shape exponents $\nu_m$, $m=1,2$.
In (d), $\tilde{u}_{2}$ indicates the peak amplitude of a fundamental nonlinear
Schrödinger soliton with wavenumber $\kappa_2$.
\label{fig:04}}
\end{figure}

\section{Two-color soliton molecules}
\label{sec:molecules}

\paragraph{Seeding of tightly bound two-color pulse compounds}
When considering initial conditions of the form of Eq.~(\ref{eq:A1A2ansatz}),
with $A_1$ a fundamental nonlinear Schrödinger soliton as in
Eq.~(\ref{eq:Sansatz}), and $A_2$ a trapped state as in Eq.~(\ref{eq:TRansatz})
with \mbox{$\phi(t)=r \,\sqrt{P_0}\,\phi_0(t)$}, the XPM contribution of the weak
trapped pulse onto the trapping soliton can be heightened by increasing the
parameter $r$.
This is demonstrated in Figs.~\ref{fig:04}(a-c), where pulse propagation
simulations in terms of the modified NSE~(\ref{eq:NSE}) are shown for different
values of $r$, significantly larger than those considered in the preceding
section.
%
Especially for larger values of $r$ [Figs.~\ref{fig:04}(b,c)], the intensity
exhibits the following dynamics: the mutual confining action of XPM results in
a contraction of both subpulses, prompting the formation of a narrow localized
pulse compound. 
A similar effect has previously been suggested by Hasegawa for multimode
optical pulses in glass fibers in Ref.~\cite{Hasegawa:OL:1980}, where he writes
``[\ldots] as many modes are trapped, the peak intensity of the packet
increases quite analogously to a gravitational instability, resulting in a
further contraction of the packet.'' (Ref.~\cite{Hasegawa:OL:1980}, p.~417).
The results shown in Figs.~\ref{fig:04}(a-c) demonstrate this effect in the
context of two-color pulse compounds in nonlinear fibers or waveguides with two
zero-dispersion points.
Let us note that, for $r\approx 1$, initial conditions as pointed out above
directly generate tightly bound, mutually confined two-color pulse compounds.
They are accompanied by radiation, emanating from the localized state upon
propagation, and can exhibit internal dynamics reminiscent molecular vibrations
\cite{Melchert:PRL:2019,Melchert:SR:2021,Oreshnikov:PRA:2022,Willms:PRA:2022}.
However, such a seeding procedure generates two-color pulse compounds in a
largely uncontrolled manner.
For completeness, we have observed the formation of similar localized pulse
compounds when taking trapped state initial conditions of the form
$\phi(t) = r \,\sqrt{P_0}\, \phi_1(t)$ for large enough $r$.

\paragraph{Simultaneous solutions of the coupled equations}
We can surpass the above seeding approach by directly searching for
simultaneous solitary-wave solutions of the coupled nonlinear
Eqs.~(\ref{eq:CNSE}) beyond the linear limit discussed in
Sect.~\ref{sec:atoms}.
Substituting an ansatz for two subpulses, labeled $m=1,2$, in the form of 
\begin{align}
A_m(z,t) = U_m(t)\,e^{i (\beta_0 + \kappa_m) z}, \quad \text{with}\quad m=1,2 \label{eq:mol_ansatz}
\end{align}
into Eqs.~(\ref{eq:CNSE}), yields two coupled ordinary differential
equations (ODEs) of second order
\begin{subequations}\label{eq:ODEs}
\begin{align}
&\ddot U_1 - \frac{2}{\beta_2^{\prime}}\left[  \gamma^\prime \left( |U_1|^2 + 2 |U_2|^2\right) - \kappa_1\right] U_1 = 0, \label{eq:2ODE1} \\ 
&\ddot U_2 - \frac{2}{\beta_2^{\prime\prime}}\left[  \gamma^{\prime\prime} \left( |U_2|^2 + 2 |U_1|^2\right) - \kappa_2\right] U_2 = 0,\label{eq:2ODE2}
\end{align}
\end{subequations}
for two real-valued envelopes $U_m\equiv U_m(t)$, $m=1,2$, with dots
denoting derivatives with respect to time.
Under suitable conditions, solitary-wave solutions for the coupled nonlinear
Eqs.~(\ref{eq:ODEs}) can be specified analytically 
\cite{Haelterman:OL:1993,Silberberg:OL:1995,Afanasjev:OL:1989,Pelinovsky:PRE:2000,Melchert:OL:2021}. 
Approximate solutions based on parameterized trial functions can be found,
e.g., in terms of a variational approach \cite{Pare:PRE:1996}.
In order to obtain simultaneous solutions $U_1(t)$, and $U_2(t)$ under more
general conditions, Eqs.~(\ref{eq:ODEs}) need to be solved numerically. 
This can be achieved, e.g., by 
spectral renormalization methods \cite{Musslimani:JOSAB:2004,Ablowitz:OL:2005,Fibich:PD:2006,Lakoba:JCP:2007}, 
shooting methods \cite{Haelterman:PRE:1994,Mitchell:PRL:1997}, 
squared operator methods \cite{Yang:SIAM:2007},
conjugate gradient methods \cite{Lakoba:PD:2009,Yang:JCP:2009}, 
$z$-propagation adapted imaginary-time evolution methods \cite{Chiofalo:PRE:2000,Yang:SIAM:2008},
or Newton-type methods \cite{Dror:JO:2016}.  
Here, in order to solve for simultaneous solutions of the ODEs (\ref{eq:ODEs}),
we employ a Newton method that is based on a boundary value Runge-Kutta
algorithm \cite{Kierzenka:ACM:2001}. 
So as to systematically obtain solutions $U_1(t)$ and $U_2(t)$, we keep five of
the six parameters that enter Eqs.~(\ref{eq:ODEs}) fixed.  Therefore we set
$\beta_2^{\prime}$, $\beta_2^{\prime\prime}$, $\gamma^{\prime}$, and
$\gamma^{\prime\prime}$ to the values considered througout the preceding
section, and preset the wavenumber $\kappa_1=|\beta_2^{\prime}|(2
t_0^2)^{-1}\approx 0.0156~\mathrm{\mu m^{-1}}$ of a fundamental nonlinear
Schrödinger soliton with $t_0=8\,\mathrm{fs}$ in Eq.~(\ref{eq:2ODE1}).
We then sweep the remaining paramter $\kappa_2$ over the wavenumber range
$(0.002,0.05)~\mathrm{\mu m^{-1}}$, enclosing the value of $\kappa_1$.
%
%
We start the parameter sweep at $\kappa_2=0.05~\mathrm{\mu m^{-1}}$, which
vastly exceeds the wavenumber eigenvalue of the lowest lying trapped state
solution at $0.0382~\mathrm{\mu m^{-1}}$.  
Above this value, we expect $U_2$ to vanish, and $U_1$ to yield a fundamental
soliton $U_1(t)=\sqrt{P_0}\,{\mathrm{sech}}(t/t_0)$ with
$P_0=|\beta_2^{\prime}|(\gamma^{\prime}\,t_0^2)^{-1}$.
We set initial trial functions for $U_1$ and $U_2$ with parity similar to the
soliton and the lowest lying trapped state, and continue the obtained solutions
to smaller values of $\kappa_2$. 
The results of this paramter sweep are summarized in Figs.~\ref{fig:04}(d-f).
We find that all solutions can be parameterized in the form
$U_m(t)=U_{0,m}\,{\mathrm{sech}}^{\nu_m}(t/t_m)$, with pulse peak amplitudes
$U_{0,m}$ [Fig.~\ref{fig:04}(d)], pulse durations $t_m$ [Fig.~\ref{fig:04}(e)],
and pulse shape exponents $\nu_m$ [Fig.~\ref{fig:04}(f)], for $m=1,2$.
In agreement with the results reported in sect.~\ref{sec:trapped_states}, we
find that a weak nonzero solution $U_2$ with $t_2=8~\mathrm{fs}$ and $\nu_2
\approx 1.55$ originates at $\kappa_2 \approx 0.038~\mathrm{\mu m^{-1}}$.
For $\kappa_2<0.038~\mathrm{\mu m^{-1}}$, the peak amplitude of the subpulse
$m=1$  continuously decreases while that for $m=2$ increases.
Below $\kappa_2\approx 0.007\mathrm{\mu m^{-1}}$, subpulse $U_1$ vanishes and
$U_2$ describes a fundamental soliton with pulse shape paramter $\nu_2=1$ and
wavenumber $\kappa_2$.
To facilitate intuition, we included the amplitude of a free soliton with
wavenumber $\kappa_2$, i.e.\ peak amplitude $\tilde{U}_{0,2}=\sqrt{2
\kappa_2/\gamma^{\prime\prime}}$, in Fig.~\ref{fig:04}(d).
Let us note that the intermediate parameter range $0.007~\mathrm{\mu
m^{-1}}<\kappa_2<0.038~\mathrm{\mu m^{-1}}$ bears tightly coupled pulse
compounds, characterized by subulse amplitudes with similar peak heights, see
Fig.~\ref{fig:04}(d).

\begin{figure}[t!]
\includegraphics[width=\linewidth]{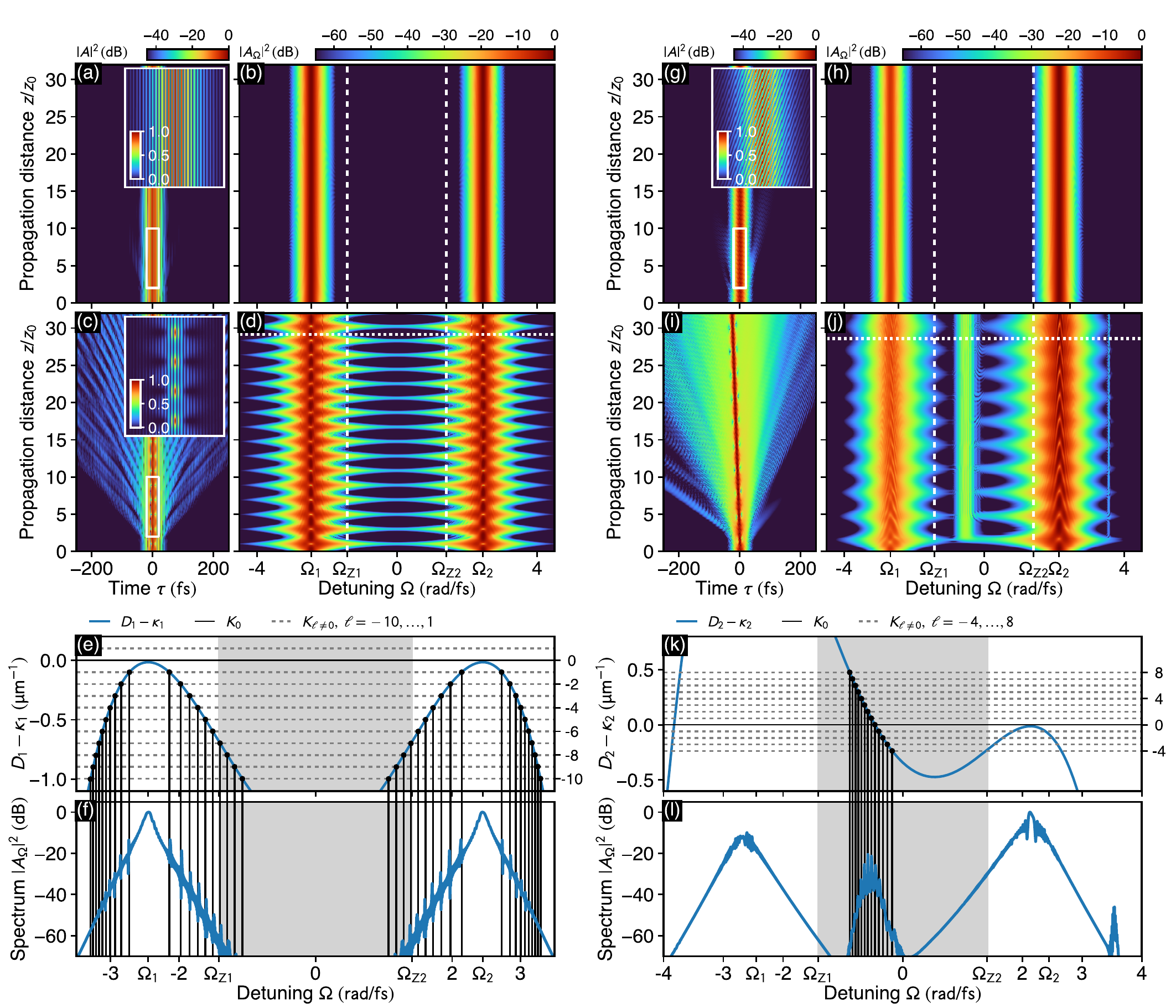}
\caption{
Resonant radiation of two-color soliton molecules.
(a,b) Stationary propagation of a soliton molecule with subpulse loci
at $\Omega_1=-\Omega_2=2.828~\mathrm{rad/fs}$.
(a) Time-domain propagation dynamics. The inset shows a close-up view of
$|A(z,\tau)|^2/{\rm{max}}(|A(0,\tau)|^2)$ in the range $\tau=-20\ldots
20~\mathrm{fs}$ and $z/z_0=2\ldots 10$.
(b) Corresponding spectrum.
(c,d) Same as (a,b) for soliton molecule order $N=1.8$.
Horizontal dashed line in (d) indicates $z/z_0=29.2$. 
(e) Dispersion profile and grapical solution of the resonance conditions Eqs.~(\ref{eq:RR})
for $z$-oscillation periods of order $m=-10\ldots 1$.
(f) Spectrum at $z/z_0=29.2$
(g-l) Same as (a-f) for a soliton molecule with subpulse loci at
$\Omega_1=-2.674~\mathrm{rad/fs}$ and $\Omega_2=2.134~\mathrm{rad/fs}$.
In (i,j) the soliton molecule order is $N=1.6$.
Horizontal dashed line in (j) indicates $z/z_0=28.6$. 
The time-domain propagation dynamics is shown in a moving frame of reference
where $\tau = t - \tilde{\beta}_1 z$. In (a,c)
$\tilde{\beta}_1=0~\mathrm{fs/\mu m}$.  In (g,i)
$\tilde{\beta}_1=0.509~\mathrm{rad/fs}$.  Propagation distance is scaled by
$z_0=32~\mathrm{\mu m}$.
Movies of the propagation dynamics shown in (a-d) and (g-j) are provided as
supplementary material under Ref.~\cite{SuppMat:Zenodo:2023}.
\label{fig:05}}
\end{figure}

\subsection{Two-color soliton pairs}

Upon closely assessing the results shown in Figs.~\ref{fig:04}(d-f), we find
that at $\kappa_2=0.0156~\mathrm{\mu m^{-1}}$, a pair of matching solutions
with plain hyperbolic-secant shape $U_m(t)=U_{0,m}\,{\mathrm{sech}}(t/t_0)$,
$m=1,2$, is attained. 
This can be traced back to the uniformity of Eqs.~(\ref{eq:2ODE1}) and
(\ref{eq:2ODE2}) for the considered set of parameters. Formally, by assuming
$\kappa\equiv \kappa_1=\kappa_2$ and $U\equiv U_1=U_2$, both equations take
the form of a standard NSE with modified paramters
\begin{align}
-\frac{\beta_2^{\prime}}{2} \frac{{\rm{d}}^2}{{\rm{d}}t^2}U(t) + 3 \gamma^{\prime} |U(t)|^2 U(t) = \kappa U(t), \label{eq:sNSE}
\end{align}
where, for convenience only, we used the parameters of Eq.~(\ref{eq:2ODE1}).
The real-valued pulse envelope $U$ should therefore be identified by the peak
intensity $\tilde{P}_0 = |\beta_2^{\prime}|(3\gamma^{\prime} t_0^2)^{-1}$, and
thus $u_1=u_2=\sqrt{1/3}\approx 0.57$ in Fig.~\ref{fig:04}(d).
Hence, at $\kappa_2=0.0156~\mathrm{\mu m^{-1}}$, both subpulses resemble 
true two-color \emph{soliton} pairs: the pulse envelopes $U_1$ and $U_2$ 
both specify a fundamental NSE soliton;
for each pulse, its binding partner modifies the nonlinear coefficient of the
underlying NSE through XPM, helping the pulse sustain its shape. 
Consequently, both pulses can only persist conjointly as a bonding unit.
This special case is consistent with a description of two-color pulse compounds
in terms of incoherently coupled pulses \cite{Melchert:OL:2021}.
By considering the ansatz Eq.~(\ref{eq:A1A2ansatz}), we can plug in the
obtained pulse envelopes for $U_1$ and $U_2$ and resubstitute the parameters
that define the propagation constant in sect.~\ref{sec:model} to obtain 
\begin{align}
A(z,t) = F(z,t) \, \cos\left( \sqrt{\frac{6 \beta_2}{|\beta_4|}} t  \right)\, e^{-i\beta_0 z},\quad \text{with}\quad F(z,t) = \sqrt{\frac{8 \beta_2}{3 \gamma t_0^2}} \, {\rm{sech}}\left(\frac{t}{t_0}\right)\,e^{i\kappa z}, \quad \text{and}\quad \kappa=\frac{\beta_2}{t_0^2}.   \label{eq:TCSM}
\end{align}
Let us note that $F$ is equivalent to the fundamental meta-soliton obtained in
Ref.~\cite{Tam:PRA:2020}, which becomes evident when substituting $\epsilon =
t_0^{-1} [3\beta_2 /(2|\beta_4|)]^{-1/2}$ and $\mu_0 \epsilon^2 =
\beta_2/t_0^2$.
This fundamental meta-soliton was first formulated by Tam \emph{et al.}, when
studying stationary solutions for the modified NSE~(\ref{eq:NSE}) by putting
emphasis on the time-domain representation of the field in terms of a
multi-scales analysis \cite{Tam:PRA:2020}. 
This unveiled a large superfamily of solitons, now referred to as generalized
dispersion Kerr solitons.
We would like to point out that within the presented approach, i.e.\ by putting
emphasis on the frequency-domain representation of two-color pulse compounds,
the fundamental meta-soliton is derived with great ease. 
Furthermore, both approaches complement each other very well. 
%
We should note that the above two-color soliton pairs resemble vector solitons
studied in the context of birefringent optical fibers
\cite{Tratnik:PRA:1988,Christodoulides:OL:1988,Kivshar:JOSAB:1990,Mesentsev:OL:1992,Tratnik:OL:1992,Afanasjev:OL:1995}.
The stationary propagation of the two-color soliton pair defined by
Eq.~(\ref{eq:TCSM}) in terms of the modified NSE~(\ref{eq:NSE}) is demonstrated
in Figs.~\ref{fig:05}(a,b).
The inset in Fig.~\ref{fig:05}(a) provides a close-up view onto the localized
pulse, indicating interference fringes with period $\Delta t \approx
\sqrt{|\beta_4|/(6\beta_2\pi^2)}\approx 1.3~\mathrm{fs}$ that are due to the
cosine in Eq.~(\ref{eq:TCSM}).
These interference fringes appear stationary since the propagation scenarion
exhibits the symmetry $\beta(\Omega_1)=\beta(\Omega_2)$ and
$\kappa_1=\kappa_2$. 
A spectrogram of the propagation scenario at $z/z_0=29.17$ is shown in
Fig.~\ref{fig:06}(a). 
A small amount of residual radiation can be seen to lie right on the curve
$\beta_1(\Omega) z$, given by the short-dashed line in Fig.~\ref{fig:06}(a). 
It was emitted by the pulse compound
during the initial propagation stage and is caused by the presence
of higher orders of dispersion at the individual subpulse loci, which were
neglected in the simplified description leading to Eqs.~(\ref{eq:TCSM}).
%


\subsection{Kushi-comb-like multi-frequency radiation}

Previously, it was shown that $z$-periodic amplitude and width oscillations of
two-color soliton molecules can be excited in a systematic manner by increasing
their initial peak amplitude by some factor $N$ according to
\mbox{$F(z,t)\leftarrow N F(z,t)$} \cite{Tam:PRA:2020,Melchert:NJP:2023}.
In analogy to usual nonlinear Schrödinger solitons, values $N>1$ define higher
order metasolitons.
Recently, we have performed a comprehensive analysis of the amplitude
oscillations of such higher order metasolitons, indicating that with increasing
$N$, the number of spatial Fourier-modes needed to characterize their periodic
peak-intensity variation, increases \cite{Melchert:NJP:2023}.
In other words, with increasing strength of perturbation of a soliton molecule,
its dynamics changes from harmonic to nonlinear oscillations. 

\paragraph{Degenerate multi-frequency radiation}
To demonstrate amplitude and width oscillations, we show the propagation
dynamics of a symmetric soliton molecule of order $N=1.8$, based on the
two-color soliton pair~(\ref{eq:TCSM}), in Figs.~\ref{fig:05}(c,d). 
As can be seen from the time-domain dynamics in Fig.~\ref{fig:05}(c), the
localized pulse exhibits periodic amplitude and width variations [close-up view
in Fig.~\ref{fig:05}(c)], and emits radiation along either direction along the
coordinate $t$ in a symmetric fashion.
Quite similar dynamics where obtained using the seeding approach in
Figs.~\ref{fig:04}(b,c).
The oscillation of the soliton molecule is also clearly visible in the 
spectrum shown in Fig.~\ref{fig:05}(d).
As evident from Fig.~\ref{fig:05}(f), at $z/z_0\approx 29.17$ it exhibits
comb-like bands of frequencies in the vicinity of the subpulse loci $\Omega_1$
and $\Omega_2$.
The location of these newly generated frequencies can be understood by
extending existing approaches for the derivation of resonance conditions
\cite{Yulin:OL:2004,Skryabin:PRE:2005,Conforti:SR:2015,Oreshnikov:PRA:2017,Melchert:SR:2020}
to two-color pulse compounds \cite{Oreshnikov:PRA:2022,Melchert:NJP:2023}.
%
%
%
Below, we summarize these resonance conditions, which where obtained by
assuming a dynamically evolving pulse compound of the form
\cite{Oreshnikov:PRA:2022} 
\begin{equation}
U_m(z,t) = \sum_{\ell} C_{m\ell}(t)\,\exp\left[ i\left( \kappa_m + K_\ell \right) z\right], \quad \text{with} \quad m\in(1,2),~\ell\in\mathbb{Z}. \label{eq:mol_vib_ansatz} 
\end{equation}
In Eq.~(\ref{eq:mol_vib_ansatz}), $C_{m\ell}$ are expansions coefficients, and 
$\kappa_m$ indicate wavenumbers that govern the $z$-propagation of each
subpulse. 
%
%
The wavenumbers of the higher harmonics of the $z$-oscillation period are
$K_\ell= 2\pi \ell/\Lambda$, with $\Lambda$ refering to the $z$-oscillation
wavelength of the pulse compound and $\ell$ labeling the corresponding order.
Based on this ansatz, the resonance conditions
\begin{subequations}\label{eq:RR}
\begin{align}
  &D_m(\Omega_{RR}) - \kappa_m = K_\ell, \quad \text{with}\quad m\in(1,2),~\ell\in\mathbb{Z},\quad \text{and} \label{eq:RR_01}\\
  &D_m(\Omega_{RR}) -2 \kappa_m + \kappa_{m^\prime} = K_\ell, \quad \text{with}\quad m, m^\prime \in (1,2),~m\neq m^\prime,~\ell\in\mathbb{Z}, \label{eq:RR_02}
\end{align}
\end{subequations}
with dispersion profiles $D_m(\Omega)\equiv \beta(\Omega)-\beta(\Omega_m)
-\beta_1(\Omega)(\Omega-\Omega_m)$ for $m=1,2$, can be derived
\cite{Oreshnikov:PRA:2022}.
In Eqs.~(\ref{eq:RR}), $\Omega_{RR}$ specifies those frequencies at which
resonant radiation (RR) is excited.
While Eq.~(\ref{eq:RR_01}) defines resonance conditions for the generation of
Cherenkov radiation by each subpulse,
Eq.~(\ref{eq:RR_02}) defines additional resonance conditions indicative of
four-wave mixing (FWM) processes involving both subpulses. 
%
%

\begin{figure}[t!]
\includegraphics[width=\linewidth]{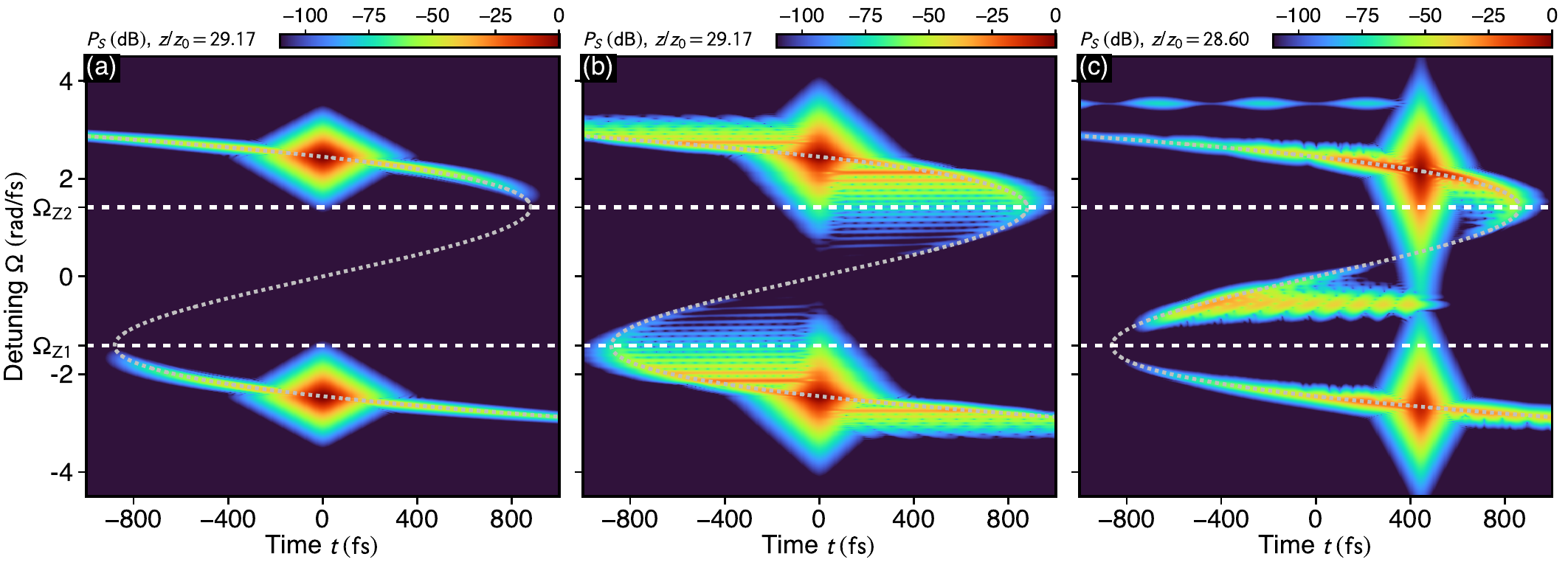}
\caption{
Spectrograms of selected soliton molecules.
(a) Two-color soliton pair of Figs.~\ref{fig:05}(a,b) at $z/z_0=29.17$, computed
using $\sigma=30~\mathrm{fs}$ in Eq.~(\ref{eq:PS}). 
(b) Oscillating, symmetric soliton molecule of Figs.~\ref{fig:05}(c,d) at
$z/z_0=29.17$ for $\sigma=30~\mathrm{fs}$, showing many narrowly spaced
resonances reminiscent of the shape of traditional Japanese Kushi combs.
(c) Oscillating, non-symmetric soliton molecule of Figs.~\ref{fig:05}(i,j) at
$z/z_0=28.6$ for $\sigma=20~\mathrm{fs}$. The pulse-wise emission of radiation,
synchronized with the periodic amplitude and width variations of the pulse
compound, is clearly visible.
In (a-c), the short-dashed line shows $\beta_1(\Omega)\, z$, indicating the
delimiting temporal position of a mode at detuning $\Omega$, emitted at $z=0$.
Movies of the propagation dynamics are provided as supplementary material under
Ref.~\cite{SuppMat:Zenodo:2023}.
\label{fig:06}}
\end{figure}

%
For the considered soliton molecule of order $N=1.8$, we find $\Lambda \approx
63~\mathrm{\mu m}\approx 2 z_0$ [with $z_0=32~\mathrm{\mu m}$, see
Fig.~\ref{fig:05}(d)].
In this case, the aforementioned symmetry $\kappa_1=\kappa_2$ renders 
Eqs.~(\ref{eq:RR_01}) and (\ref{eq:RR_02}) degenerate.
As evident from the graphical solution of Eqs.~(\ref{eq:RR}) in
Fig.~\ref{fig:05}(e), the resonance conditions predict the newly generated
frequencies in Fig.~\ref{fig:05}(f) very well.
A spectrogram of the propagation scenario at $z/z_0=29.17$ is shown in
Fig.~\ref{fig:06}(b).
Therein, the multi-peaked spectral bands, at which the oscillating soliton
molecule sheds radiation, are reminiscent of the shape of traditional Japanese
Kushi combs.

\paragraph{Non-degenerate multi-frequency radiation}

Let us note that, due to the wide variety of two-color pulse compounds with
different substructure, their emission spectra manifest in various forms.
For example, considering a pair of group-velocity matched detunings different
from the one considered above, the degeneracy among Eqs.~(\ref{eq:RR}) can be
lifted.
Subsequently we take $\Omega_1=-2.674~\mathrm{rad/fs}$ and
$\Omega_2=2.134~\mathrm{rad/fs}$, for which
$\beta_1^{\prime}=0.514~\mathrm{fs/\mu m}$,
$\beta_1^{\prime\prime}=0.514~\mathrm{fs/\mu m}$,
$\beta_2^{\prime}=-2.576~\mathrm{fs^2/\mu m}$, and,
$\beta_2^{\prime\prime}=-1.278~\mathrm{fs^2/\mu m}$.
In terms of the coupled ODEs~(\ref{eq:ODEs}) we then determine a pair
of simultaneous solutions which specify the initial condition
\begin{align}
A_0(t) = U_{0,1}\,{\rm{sech}}^{\nu_1}\left( \frac{t}{t_1} \right) \, e^{-i\Omega_1 t}
       + U_{0,2}\,{\rm{sech}}^{\nu_2}\left( \frac{t}{t_2} \right) \, e^{-i\Omega_2 t}, \label{eq:mol2}
\end{align}
with parameters
$U_{0,1} = 0.050~\mathrm{\sqrt{W}}$,
$U_{0,2} = 0.141~\mathrm{\sqrt{W}}$,
$t_1 = 7.207~\mathrm{fs}$,
$t_2 = 7.271~\mathrm{fs}$,
$\nu_1 = 0.901$, and
$\nu_2 = 1.022$.
The stationary propagation of this soliton molecule with non-identical 
subpulses is shown in Fig.~\ref{fig:04}(g,h).
As a consequence of the broken subpulse-symmetry, the interference fringes that
characterize the pulse compound are not stationary any more [close-up view in
Fig.~\ref{fig:05}(g)].
The fact that the pulse compound remains localized, despite its envelope
exhibiting a non-stationary profile, might be the reason why no such objects
could be found using a time-domain based Newton conjugate-gradient method
\cite{Tam:PRA:2020}.
Next, we increase the order of this soliton molecule to $N=1.6$, resulting in
the propagation dynamics with $z$-oscillation period $\Lambda \approx
106~\mathrm{\mu m} \approx 3.3 z_0$ shown in Figs.~\ref{fig:05}(i,j).
In this case, a pronounced mulit-peaked spectral band of frequencies within 
the domain of normal dispersion is excited [see Figs.~\ref{fig:05}(j,l)].
These newly generated frequencies can be linked to multi-frequency Cherenkov
radiation emitted by the subpulse at $\Omega_2$, as can be seen from the
graphical solution of the resonance conditions~(\ref{eq:RR_01}), shown in
Fig.~\ref{fig:05}(k).
Let us note that similar coupling phenomena of localized states to the
continuum have have earlier been observed for solitons in periodic dispersion
profiles \cite{Conforti:SR:2015}, oscillating bound solitons in twin-core
fibers \cite{Oreshnikov:PRA:2017}, and dissipative solitons in nonlinear
microring resonators \cite{Melchert:SR:2020}.
%
A further band of frequencies, excited in the vicinity of $\Omega\approx
3.5~\mathrm{rad/fs}$ can be attributed to FWM-resonances described by
Eq.~(\ref{eq:RR_02}).
A spectrogram of the propagation scenario at $z/z_0=28.6$ is shown in
Fig.~\ref{fig:06}(c), unveiling that the resonant radiation emanates from the
oscillating soliton molecule in a pulse-wise fashion.

%

\section{Summary and conclusions}
\label{sec:conclusion}

In summary, we have discussed several aspects of the $z$-propagation of
two-color pulse compounds in a modified NSE with positive group-velocity
dispersion coefficient and negative fourth-order dispersion coefficient.
Therefore, we considered the interaction dynamics of two pulses in distinct
domains of anomalous dispersion, group-velocity matched despite a large
frequency gap.

We have demonstrated that their mutual confining action can manifest itself in
different forms, depending on the relative strength of SPM and XPM felt by each
pulse. 
In the limiting case where the resulting bound states consist of a strong
trapping pulse, given by a soliton, and a weak trapped pulse, we have shown
that optical analogues of quantum mechanical bound states can be realized that
are determined by a Schrödinger-type eigenvalue problem
\cite{Melchert:PRL:2019}.
The resulting photonic meta-atoms even support Rabi-type oscillations of its
trapped states, similar to the recurrence dynamics of wave packets in quantum
wells \cite{Bocchieri:PR:1957}.
We further probed the limits of stability of these meta-atoms by imposing a 
group-velocity mismatch between the trapping soliton and the trapped pulse.
With increasing strength of perturbation, parts of the trapped state escapes
the soliton, similar in effect to the ionization of quantum mechanical atoms.
These findings complement our earlier results on the break-up dynamics of
two-color pulse compounds \cite{Melchert:SR:2020}. 

For the more general case where the mutual confining action between the pulses
is dominated by XPM, we have discussed a simplified modeling approach, allowing
to determine simultaneous solutions for the bound pair of pulses.
The resulting solutions feature the above meta-atoms as limiting cases when 
the disparity of the subpulse amplitudes is large.
Further, by exploiting symmetries of the underlying propagation model, a
special class of solutions, forming true two-color soliton pairs
\cite{Melchert:OL:2021}, was characterized in closed form.
This special class of solutions, referred to as generalized dispersion Kerr
solitons, has also been derived in Ref.~\cite{Tam:PRA:2020}. 
We have presented numerical results demonstrating the complex propagation
dynamics of such pulse compounds, which we here referred to as two-color
soliton molecules.
Specifically, we have shown that soliton molecules exhibit highly robust 
vibrational characteristics, a behavior that is difficult to achieve in 
a conservative NSE system.
These non-stationary, $z$-periodic dynamics of the subpulses triggers the
emission of resonant radiation. 
The location of the resulting multi-peaked spectral bands can be precisely
predicted by means of phase-matching conditions
\cite{Oreshnikov:PRA:2022,Melchert:NJP:2023}.
Due to the manifold of soliton molecules with different substructure, their
emission spectra manifest in various complex forms.
Most notably, if the oscillating soliton molecule consists of a pair of
identical subpulses, inherent symmetries lead to degeneracies in the resonance
spectrum, causing their spectrogram trace to resemble the shape of Japanese
Kushi combs.
Additional perturbations lift existing degeneracies and result in more complex
emission spectra which are characterized by distinct spectral bands that can be
separately linked to resonant Cherenkov radiation and additional four-wave
mixing processes.
The occurrence of such multi-frequency radiation, especially in the degenerate
form, comprises a fundamental phenomenon in nonlinear waveguides with multiple
zero-dispersion points and sheds light onto the puzzling propagation dynamics
of two-frequency pulse compounds, resembling the generation of radiation by
vibrating molecules.

%

Finally, let us note that we recently extended the range of systems in which 
such two-color pulse compounds are expected to exist.
Therefore, we considered waveguides with a single zero-dispersion point and
frequency dependent nolinearity with a zero-nonlinearity point
\cite{Driben:OE:2009,Bose:JOSAB:2016}.
In such waveguides, soliton dynamics in a domain of normal dispersion can be
achieved by a negative nonlinearity \cite{Bose:PRA:2016,Arteaga:PRA:2018}.
In the corresponding description of pulse compounds in terms of the simplified
model~(\ref{eq:CNSE}), having $\beta_2^{\prime}<0$ and
$\beta_2^{\prime\prime}>0$ then requires $\gamma^{\prime}>0$ and
$\gamma^{\prime\prime}<0$, and the potential well in the eigenproblem
corresponding to Eq.~(\ref{eq:EVproblem}) is ensured by
$\gamma^{\prime\prime}<0$ \cite{Melchert:OL:2023}.
We studied the above binding mechanism for incoherently coupled two-color pulse
compounds in such waveguides, demonstrating meta-atoms and molecule-like bound
states of pulses that persist in the presence of the Raman effect
\cite{Willms:PRA:2022,Melchert:OL:2023}, allowing to understand the complex
propagation dynamics observed in a recent study on higher-order soliton
evolution in a photonic crystal fiber with one zero-dispersion point and
frequency dependent nonlinearity \cite{Zhao:PRA:2022}.

\section*{Acknowledgements}

We acknowledge support from the Deutsche Forschungsgemeinschaft  (DFG) under
Germany’s Excellence Strategy within the Cluster of Excellence PhoenixD
(Photonics, Optics, and Engineering – Innovation Across Disciplines) (EXC 2122,
projectID 390833453).


\bibliographystyle{elsarticle-num} 
\bibliography{references.bib}





\end{document}